# Cascading Sensor Network Clock Synchronization Scheme

Meng ZHOU, Jie WU, Member IEEE, Xuesong LIU

*Abstract*–In this paper, we propose a hybrid clock synchronization architecture for a cascaded sensor network based on GPS time service and synchronous frame protocol. The sensor's upper-level unit is called sensor management unit (SMU) which synchronizes the local clock with the GPS and then use the broadcast sync frame to synchronize the clock of the sensor chain. In this paper, we build a link of 20 cascaded sensors to measure the center frequency and clock stability. The results showed that the center frequency deviation was 0.037 Hz and the stability was 0.045 ppm to 0.066 ppm .The stability of GPS is 0.06ppm, the scheme has achieved the effect similar to installing GPS chip on every sensor.

*Index Terms*—cascading sensor network, clock Synchronization

## I. INTRODUCTION

SENSOR network emerged as a promising area of research in recent years and have been widely used in applications such as data acquisition, environmental monitoring and industrial automation. In many applications involving a large number of sensors, global clock synchronization is an integral part of the sensor network. In general, clocks run at slightly different frequencies due to the imperfections in quartz crystal and environmental conditions, such as initial manufacturing tolerance, aging, humidity, pressure, temperature, and other environmental factors. Local clock drift can cause clock discrepancies between sensors over time, affecting some periodic operation of the sensor network. So sensor network must have a global clock synchronization mechanism to reduce the sensor clock drift. However, GPS-based synchronization is not suitable for low-power applications. It is not possible for all the sensors to be installed with a GPS chip for clock synchronization. In addition, the actual cascaded sensor network has a large amount of data transmission on the uplink from the sensor to the SMU, the transmission delay is not fixed due to the frame buffer mechanism. Therefore, the precision time protocol is also not suitable, the sync frame can only appear in the downlink channel from SMU to sensors, and sensors does not need to respond the sync frames.

Manuscript received February 1, 2018. This project supported by NSFC (Grant No. 41574106), Major National Science and Technology Special Program of China (Grant No. 2017ZX05008-008-041), R&D of Key Instruments and Technologies for Deep Resources Prospecting (Grant No. ZDYZ2012-1-05-03).

The authors are with the Department of Modern Physics and State Key Laboratory of Particle Detection and Electronics, University of Science and Technology of China, Hefei, Anhui 230026, China (e-mail: wujie@ustc.edu.cn).

## II. SMU LOCAL CLOCK SYNCHRONIZATION.

SMU through the control of the VCXO voltage control MCU peripheral timer, the use of the timer capture function to count the second pulse interval. Adjust the VCXO voltage by comparing the difference between the count value and 32768000HZ, adjust the frequency to 32.768MHz to achieve local clock frequency synchronization with GPS .

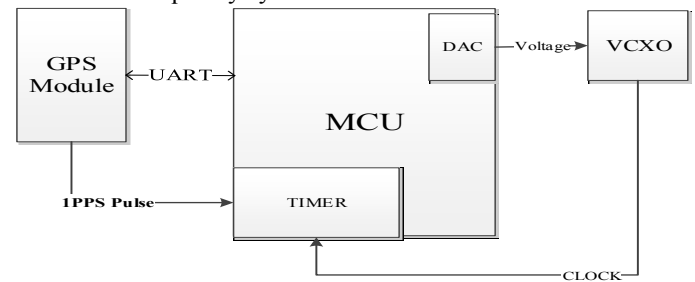

Figure.1. SMU synchronize local clock with GPS

.Whenever SMU receives a second pulse from GPS, the MCU will generate a synchronization frame. The MCU sends the frame to the FPGA. When the synchronization frame is sent from the FPGA, the FPGA adds the latched value of the time counter to the frame and sends it to the lower sensor chain as shown in figure.2

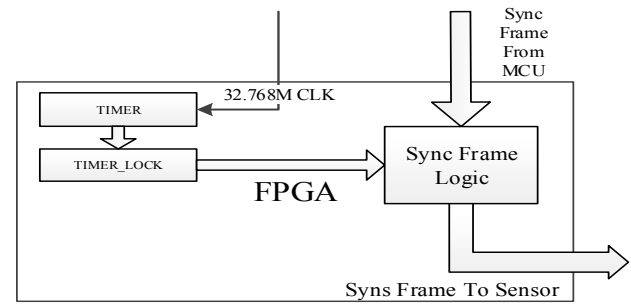

Figure.2. SMU to generate synchronous frame

## III. CLOCK SYNCHRONIZATION OF SENSOR AND SMU

The sensor receives the synchronization frame sent by SMU and does not respond. The sensor changes the SMU count value in the synchronization frame to a local count value, which is then forwarded to the next sensor so that the clock frequency of the entire cascading sensor chain is synchronized, sensor and SMU synchronization process shown in Figure.3. Delay is the transmission delay between two adjacent stations, Offset is deviation between two adjacent stations. The synchronization frame contains the count value of the previous station when it is sent. When the sensor receives the synchronization frame, the latched local count value can be obtained. Since T2-T1 and T2`-T1` are equal, by adjusting the voltage value of the local VCXO, the sensor realizes that the difference between the local count value and the previous station count value is the same. Realize the local clock frequency and the previous station clock frequency synchronization.

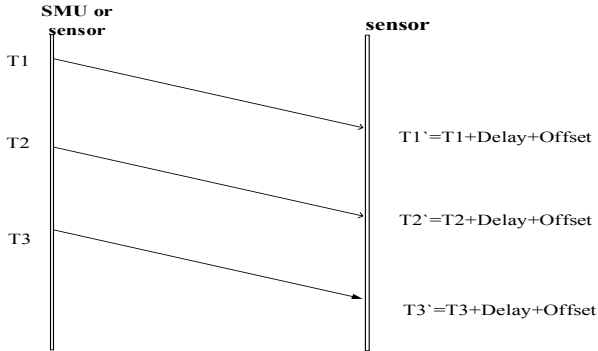

Figure.3. Clock synchronization of sensor and SMU

## IV. TEST

Test structure shown in Figure.4. Owing to we only have 20 sensors, the synchronization results of 20 sensors need to be measured, and the results are linearly fitted to infer the synchronization status of the 20th station. Sensor and SMU, sensor and sensor each use a 1-meter cable connection. Sensor outputs the clock after synchronization, the signal input to the frequency meter to measure. Twenty sensors to take interval measurements, measuring 2, 4, 6, 8, 10, 12, 14, 16, 18, 20 ten sensors. Each sensor recorded 600 data to calculate the stability and center frequency of the clock.

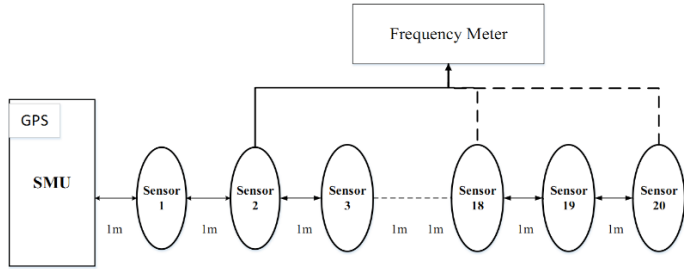

Figure.4. Test cascade topology

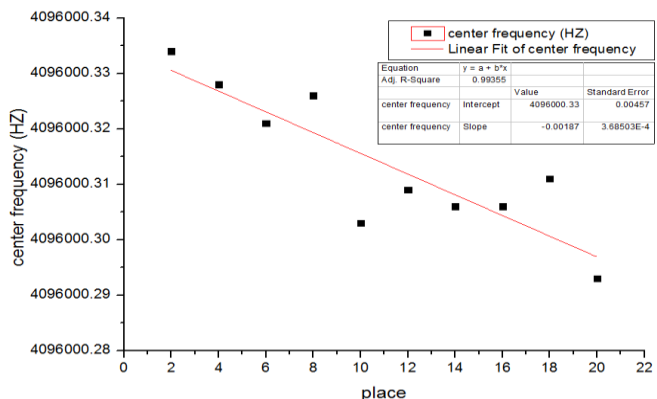

Figure.5. clock center frequency fitting results

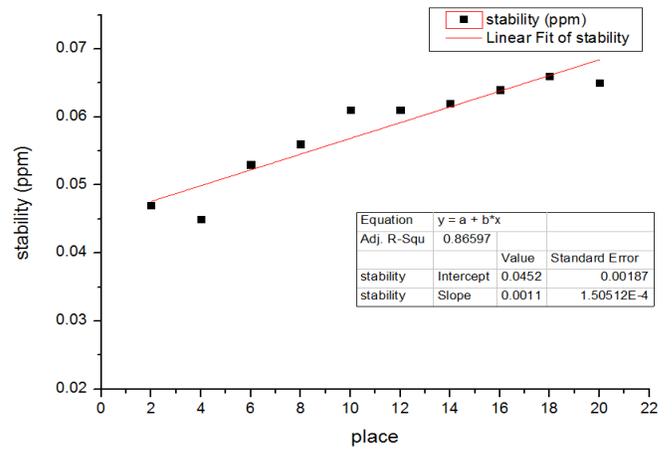

Figure.6. Clock frequency stability fitting results

According to the results shown in Fig.5 and Fig.6, the slope of the fitted center frequency straight line is -0.00187 and the stability straight line slope is 0.0011. Calculations show that the center frequency deviation of the sensors chain is about 0.037Hz and the stability deviation is 0.022ppm.

## V. CONCLUSION

In this paper, we propose a hybrid clock synchronization architecture to realize the clock synchronization of cascaded sensor networks, which is of practical value for the cascaded links that are suitable for one-way synchronization frames. The stability of GPS is 0.06ppm, the scheme has achieved the effect similar to installing GPS chip on every sensor. As we have reached our actual needs, we have not continued to improve the scheme. If the pursuit of better clock frequency synchronization accuracy and stability, you can try to improve the VCXO regulation accuracy and more high-frequency clock instead of 32.768M clock.